\documentclass{article}

\usepackage{arxiv}

\usepackage[utf8]{inputenc} 
\usepackage[T1]{fontenc}    
\usepackage{graphicx}

\title{Towards a Passive BCI to Induce Lucid Dream}

\author{
  Morgane~Hamon \\
  Ullo\\
  La Rochelle, France\\
  \texttt{morgane@ullo.fr} \\
   \And
  Emma~Chabani \\
  Institut du Cerveau et de la Moelle épinière\\
  Paris, France\\
  \texttt{emma.chabani@gmail.com} \\
  \And
  Philippe~Giraudeau \\
  Inria\\
  Bordeaux, France\\
  \texttt{philippe.giraudeau@inria.fr} \\
}

\begin{document}
\maketitle

\begin{abstract}
Lucid dreaming (LD) is a phenomenon during which the person is aware that he/she dreaming and is able to control the dream content. Studies have shown that only 20\% of people can experience lucid dreams on a regular basis. However, LD frequency can be increased through induction techniques. External stimulation technique relies on the ability to integrate external information into the dream content. The aim is to remind the sleeper that she/he is dreaming. If this type of protocol is not fully efficient, it demonstrates how sensorial stimuli can be easily incorporated into people's dreams. The objective of our project was to replicate this induction technique using material less expensive and more portable. This material could simplify experimental procedures. Participants could bring the material home, then have a more ecological setting. First, we used the OpenBCI Cyton, a low-cost EEG signal acquisition board in order to record and manually live-score sleep. Then, we designed a mask containing two LEDs, connected to a microcontroller to flash visual stimulation during sleep. We asked two volunteers to sleep for 2 hours in a dedicated room. One of the participants declared having a dream during which the blue lights diffused by the mask were embedded into the dream environment. The other participant woke up during the visual stimulation. These results are congruent with previous studies. This work marked the first step of a larger project. Our ongoing research includes the use of an online sleep stage scoring tool and the possibility to automatically send stimuli according to the sleep stage. We will also investigate other types of stimulus induction in the future such as vibro-tactile stimulation that showed great potentials.

\end{abstract}

\keywords{Lucid Dream \and Induction techniques \and Sleep \and Visual stimulation \and EEG}

\section{Introduction}
During a normal night’s sleep, people enter on a cyclical basis different vigilance states: Wakefulness, N1, N2, N3 and REM sleep \cite{silber2007visual}. A sleep cycle lasts around 90 minutes, and occurs 4 to 6 times a night. The N1, N2 and N3 sleep stages are usually grouped as N-REM (Non-REM) as an opposition to the REM sleep. It was long believed that dreams exclusively occur during REM sleep \cite{dement1957relation} but this hypothesis has been refuted \cite{foulkes1962dream}. Awakenings during N-REM sleep showed dreams recall as well \cite{nielsen2001rem}. Lucid dreaming (LD) is a phenomenon during which the person is aware that he/she dreaming and is able to control the dream content \cite{laberge1985lucid}. 
LD can be used to enhance the dream content, train physical skills \cite{erlacher2017sleep} or avoid nightmares. That’s why authors have studied the ability to induce LD or increase the LD frequency \cite{gackenbach1988lucid}. 
Thus, this project aims to reproduce induction techniques to help people experience LD. As EEG amplifiers became less expensive and more portable, we aim at proposing a solution that could be deployed at home.

\section{Visual induction to Augment LD}
This pilot study has been conducted during a hackathon, a 48-hours event that occurred in December 2017 \footnote{https://mindlabdx.github.io/hack1cerveau/}. We used a electroencephalogram (EEG) headset to acquire sleep data like sleep stages. First, we built a EEG cap out of swim cap on which electrodes are sewed (see Fig.\ref{fig:fig1}, \textit{Left}). They are connected to the OpenBCI Cyton Board, that is sending data to a computer via Bluetooth. The OpenBCI software allows to visualize brainwaves obtained from the electrodes. Sleep stages, arousal, eye movement were deducted from the neural oscillations and were scored according to international criteria \cite{rechstchaffen1968manual}.
Finally, a mask has been created with two LED placed above each eye to send flash light (see Fig.\ref{fig:fig1}, \textit{Right}). Lights were controlled by an Arduino Uno connected to the same computer which runs the openBCI software. We sent visual stimulation when the participant enters in REM sleep. REM stage for Rapid Eye Movements has an EEG activity mainly branded by theta waves (5-7 Hz). There is a complete muscle atonia.

\begin{figure}
  \centering
  \includegraphics[scale=0.5]{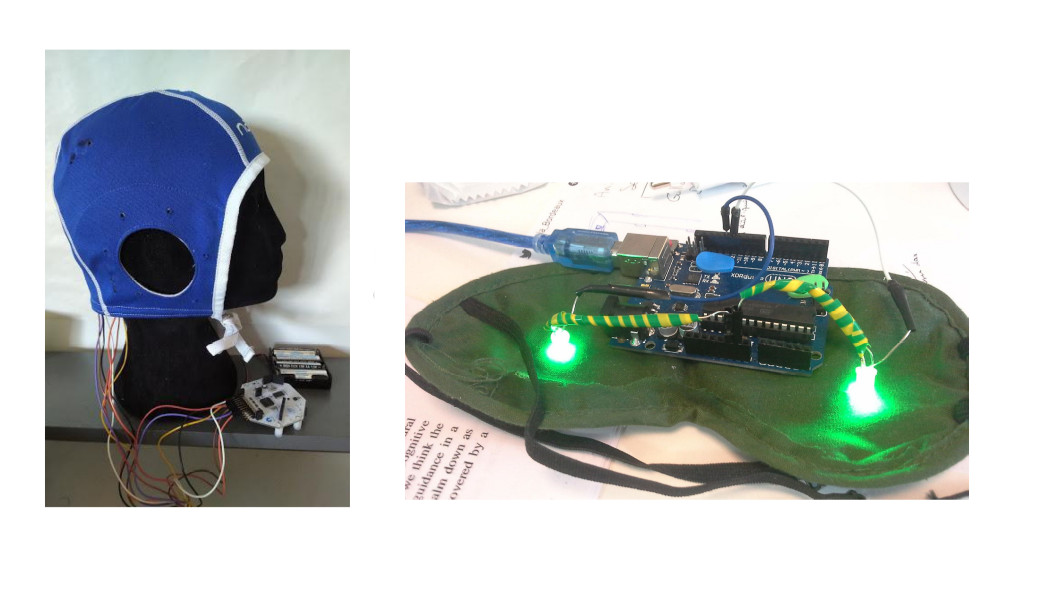}
  \caption{\textit{Left:} EEG cap with electrodes connected to an OpenBCI Cyton, \textit{Right:} The mask with LEDs and the micro-controller (Arduino UNO)}
  \label{fig}
\end{figure}

\section{Pilot Study}

Two volunteers (1 male 24y, 1 female, 26y), were not naive about lucid dreaming and had already experienced it. They were also considered as recallers, who are subjects with high Dream Recall Frequency (DRF): remembering dreams more than two nights a week \cite{cory1975predicting}.          
The volunteers were equipped with our EEG headset, our mask and earplugs to help them fall asleep. Subsequently visual stimulations were presented to determine the intensity of the LEDs. Then participants were told they could sleep for about 2 hours. 
As the system wasn’t live-scoring sleep, sleep stages were monitored by our team and stimuli were sent when needed (REM sleep). The method is pictured in Fig.\ref{fig}. 

\begin{figure}
  \centering
  \includegraphics[scale=0.2]{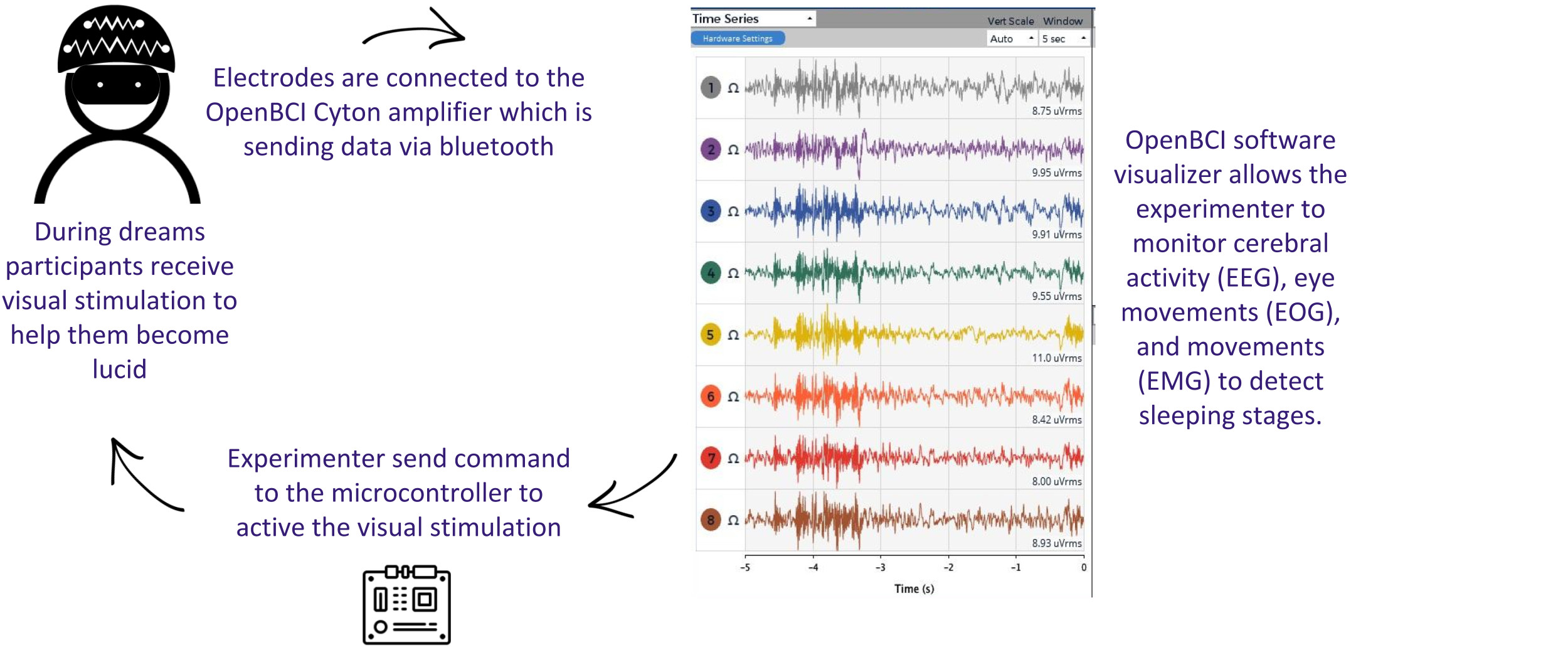}
  \caption{Cerebral activity, eye movements and movements are monitored and visualized in real-time to determine the sleep stage the volunteer is going through. Visual stimulations are sent when the participant reaches REM sleep. }
  \label{fig:fig1}
\end{figure}

\section{Results}

We observed that the blue flashing light was integrated in dream content for both participants. One participant reported to be behind a fish tank window and saw blue flashes wondering if it was natural or not but without understand it was a signal. However it did not trigger lucid dreaming. The second participant woke up right after the first stimulation. These results are congruent with previous studies \cite{laberge1985lucid}, \cite{paul2014lucid}.

\section{Discussion and Future plan}

This pilot study allowed us to better know how and when sending stimuli, an important step towards a passive BCI \cite{zander2011towards}. Our ongoing research will explore different modalities of visual stimulation such as the color, the intensity and the frequency in order to determine the best combination to induce LD. Vibro-tactile stimulation showed great potentials as well \cite{stumbrys2012induction}. We will also compare different devices to detect sleep stages (an EEG headband with less electrodes and an actimeter). This will be the first step to build a device able to automatically discriminate sleep stages.

\bibliographystyle{apalike}
\bibliography{references}

\end{document}